\documentclass[aps,prb,reprint,showpacs,floatfix]{revtex4-1}
\usepackage{amsmath,graphicx,color,verbatim}

\begin{document}

\title{A materials informatics approach to the identification of one-band correlated materials analogous to the cuprates}
\author{Eric B. Isaacs}
\author{Chris Wolverton}
\email{c-wolverton@northwestern.edu}
\affiliation{Department of Materials Science and Engineering, Northwestern University, Evanston, Illinois 60208, USA}

\begin{abstract}
One important yet exceedingly rare property of the cuprate
high-temperature superconductors is the presence of a single
correlated $d$ band in the low-energy spectrum, leading to the
one-band Hubbard model as the minimal description. In order to search
for materials with interesting strong correlation physics as well as
possible benchmark systems for the one-band Hubbard model, here we
present a new approach to find one-band correlated materials analogous
to the cuprates by leveraging the emerging area of materials
informatics. Using the composition, structure, and formation energy of
more than half a million real and hypothetical inorganic crystalline
materials in the Open Quantum Materials Database, we search for
synthesizable materials whose nominal transition metal $d$ electron
count and crystal field are compatible with achieving an isolated
half-filled $d$ band. Five Cu compounds, including bromide, oxide,
selenate, and pyrophosphate chemistries, are shown to successfully
achieve the one-band electronic structure based on density functional
theory band structure calculations. Further calculations including
magnetism and explicit on-site Coulomb interaction reveal significant
evidence for strong correlation physics in the five candidates,
including Mott insulating behavior and antiferromagnetism. The success
of our data-driven approach to discovering new correlated materials
opens up new avenues to design and discover materials with rare
electronic properties.
\end{abstract}

\date{\today} \maketitle

\section{Introduction}\label{sec:intro}

The cuprates, one of the most famous classes of materials in condensed
matter physics, are layered copper-oxide ceramics with copper-oxygen
planes exhibiting unconventional, high-temperature
superconductivity.\cite{bednorz_possible_1986} Despite decades of
study, the details of the mechanism and phase diagram of this class of
materials is still not fully settled, due in part to a complex phase
diagram in which doping the antiferromagnetic Mott insulating parent
compound can lead to pseudogap, charge density wave, spin density
wave, and ``strange metal'' phases in addition to the superconducting
phase. However, it is generally accepted that the presence of a
\textit{single d orbital} at low energy is a very important
characteristic leading to the high critical temperature $T_c$ for
superconductivity in the
cuprates.\cite{lee_doping_2006,pickett_electronic_1989}

A minimal model for the physics of the cuprates is the one-band
Hubbard model (1BHM) corresponding to the
Hamiltonian $$\hat{H}_{1BHM}=-t\sum_{i,j,\sigma}\hat{c}_{i\sigma}^{\dagger}\hat{c}_{j\sigma}+U\sum_{
  i}\hat{n}_{i\uparrow}\hat{n}_{i\downarrow}.$$ Here
$\hat{c}_{i\sigma}^{(\dagger)}$ annihilates (creates) an electron of
spin projection $\sigma$ on lattice site $i$,
$\hat{n}_{i\sigma}=\hat{c}_{i\sigma}^{\dagger}\hat{c}_{i\sigma}$, $t$
is the hopping parameter, and the on-site Coulomb repulsion $U$
parametrizes the strength of the electronic correlations. Although the
1BHM is easy to write down, this many-body Hamiltonian has only been
solved in one (using the Bethe ansatz\cite{lieb_absence_1968}) and
infinite (using dynamical mean-field
theory\cite{georges_hubbard_1992}) dimensions. In general, $t$ leads
to delocalized electronic states and metallic behavior, whereas $U$
localizes the electrons. In correlated materials like the parent
compounds of the cuprates, $U$ is large with respect to $t$ and (Mott)
insulating behavior is found.


Thus, the cuprates can be considered \textit{correlated one-band
  materials}. We are not aware of any correlated one-band material
outside of the cuprates. One key reason for their rarity is that most
transition metal compounds have octahedral or tetrahedral
coordination, for which the 5 $d$ orbitals split into a group of 3
($T_{2g}$ for octahedral) and a group of 2 ($E_g$ for octahedral)
according to group theory. Therefore, while a multiband Hubbard model
is considered to be achieved in many compounds, realizations of the
1BHM are very rare. The discovery of such materials would be highly
desirable for two purposes:
\begin{enumerate}
\item To search for new unconventional superconductors or materials
  exhibiting other interesting strong correlation phenomena, and
\item To provide physical realizations of the 1BHM to serve as
  benchmarks for our theories of strong correlation physics.
\end{enumerate}

Recent efforts to achieve such a material have not been successful.
For example, Chaloupka and Khaliullin proposed to break the symmetry
of the $E_g$ levels in octahedral LaNiO$_3$ by forming a superlattice
with LaAlO$_3$.\cite{chaloupka_orbital_2008} Unfortunately, the
experiments suggest that this approach yields insufficient symmetry
breaking to achieve a one-band model.\cite{wu_strain_2013} LiCuF$_3$
and related compounds were proposed in the trigonal bipyramidal
coordination, whose symmetry does lead to a singly-degenerate
level.\cite{griffin_bespoke_2016} Although it was shown to have
promising characteristics, LiCuF$_3$ is not thermodynamically stable
under ambient conditions.\cite{griffin_bespoke_2016}



Here we employ the emerging area of materials
informatics\cite{curtarolo_high-throughput_2013,agrawal_perspective_2016,jain_new_2016}
to accelerate the search for one-band correlated materials analogous
to the cuprates. In order to identify candidate materials, we devise a
materials database search strategy based on (1) the elements contained
in the material, (2) the local coordination geometry of atoms in the
crystal structure, (3) nominal valence electron count for each
element, and (4) thermodynamic stability. In particular, we look for
(1) compounds containing transition metals and anions, (2) transition
metals coordinated by anions in a coordination environment whose
crystal field leads to a singly-degenerate $d$ level, (3) transition
metals with a nominal $d$ electron count leading to half filling of
such a level, and (4) compounds that are thermodynamically stable or
nearly stable. For example, the cuprate parent compound La$_2$CuO$_4$
would satisfy criteria (2) and (3) since Cu is in a square planar
environment and nominally in a $d^9$ configuration, half-filling a
singly-degenerate $B_{1g}$ ($d_{x^2-y^2}$) level. We query for any
crystal that simultaneously satisfies all four criteria. We execute
our strategy using the Open Quantum Materials Database (OQMD), an
extensive electronic structure database with calculations for over
550,000 real and hypothetical inorganic crystalline materials (as of
June 2017).

We successfully identify five correlated one-band materials: CuBr$_2$,
Li$_2$CuO$_2$, BaCu(SeO$_3$)$_2$, SrCu(SeO$_3$)$_2$, and
K$_3$H(CuP$_2$O$_7$)$_2$. DFT calculations incorporating corrections
for electronic correlations demonstrate promising characteristics in
these materials including antiferromagnetism and Mott insulating
behavior. Additionally, our findings illustrate the power of
high-throughput computing and materials informatics to search for
complex materials possessing rare electronic properties.

\section{Methodology}\label{sec:methods}

\subsection{Screening strategy}

Our materials informatics screening approach consists of several
materials properties that must be simultaneously satisfied:
\begin{enumerate}
  \item \textbf{Chemistry} -- Compound must contain transition metal
    and anion elements, and (for practicality) no radioactive elements
  \item \textbf{Crystal structure} -- In the compound, transition
    metals (TMs) must be in a coordination whose local symmetry yields
    one or more singly-degenerate $d$ orbitals based on crystal field
    theory. The coordinations considered are linear, trigonal planar,
    square planar, trigonal bipyramidal, square pyramidal, trigonal
    prismatic, pentagonal bipyramidal, and square antiprismatic.
  \item \textbf{Electron count} -- All TM in the compound must have a
    nominal $d$ electron count corresponding to a half-filled
    singly-degenerate level given the crystal field splitting of the
    coordination environment.
  \item \textbf{Thermodynamics} -- The compound must be no more than
    25 meV/atom above the thermodynamic ground state (as determined
    via convex hull analysis) and/or reported experimentally. This
    criterion is designed to focus on compounds which are likely to be
    synthesizable. The particular threshold value of 25 meV/atom is
    chosen to match the magnitude of computed hull distances for
    synthesized metastable compounds found by Sun \textit{et al.} for
    most chemistries.\cite{sun_thermodynamic_2016}
\end{enumerate}

Our search strategy is executed on the OQMD,
\cite{saal_materials_2013,kirklin_open_2015} an open database
containing calculations of over half a million known and hypothetical
inorganic crystalline compounds derived from the Inorganic Crystal
Structure Database
(ICSD)\cite{bergerhoff_inorganic_1983,belsky_new_2002} and structural
prototypes. The OQMD contains electronic structure calculations at the
DFT\cite{hohenberg_inhomogeneous_1964,kohn_self-consistent_1965} and
in some cases DFT+$U$ level at consistent sets of parameters to enable
consistent thermodynamic analysis.

We include details on the coordination environments considered in this
study in the Supplemental Material.\footnote{See Supplemental Material
  for details on the coordination environments, false positive
  examples, and the complete list of 187 candidate compounds.} A
description of our method for nominal electron counting and for the
local structural queries performed to ascertain coordination
environments are included in our previous
work.\cite{isaacs_inverse_2018} The coordination environment query
relies on separate computation (outside of the qmpy
framework\cite{qmpy}) of the TM nearest-neighbor bond lengths and
angles, which are not directly stored in the OQMD.

Our screening strategy only requires computation of the electronic
band structure as a post-processing step for a small number of
candidate materials. An alternative approach of directly analyzing the
electronic band structures that exist in materials
databases\cite{curtarolo_aflowlib_2012,jain_commentary_2013} is not
applicable for our purpose primarily because such databases only
contain band structures already including spin polarization and
Hubbard $U$ effects, rather than the underlying non-spin-polarized DFT
band structure. This is problematic since the desired isolated band,
if present, can be hidden once it is spin-split and pushed into other
bands, as in the case of VS$_2$.\cite{isaacs_electronic_2016}

\subsection{Calculation details}

For the most promising materials identified, we perform additional DFT
and DFT+$U$ calculations to investigate the electronic and magnetic
properties in more detail. We employ the Vienna \textit{ab initio}
simulation package
(\textsc{vasp})\cite{kresse_ab_1994,kresse_ab_1993,kresse_efficient_1996,kresse_efficiency_1996}
to perform generalized gradient
approximation\cite{perdew_generalized_1996} calculations and employ
the rotationally-invariant Hubbard $U$
interaction.\cite{liechtenstein_density-functional_1995} We use the
projector augmented wave method,
\cite{blochl_projector_1994,kresse_ultrasoft_1999} a 600 eV kinetic
energy cutoff, and uniform $k$-point meshes of $k$-point density of
350/\AA$^{-3}$ or greater. The ionic forces and total energy are
converged to 0.001 eV/\AA\ and 10$^{-6}$ eV, respectively. The
high-symmetry $k$-point paths for band structures are based on the
AFLOW conventions.\cite{setyawan_high-throughput_2010}

\section{Results and Discussion}\label{sec:results}

\subsection{High-throughput materials screening}

\begin{figure}[htbp]
  \begin{center}
    \includegraphics[width=1.0\linewidth]{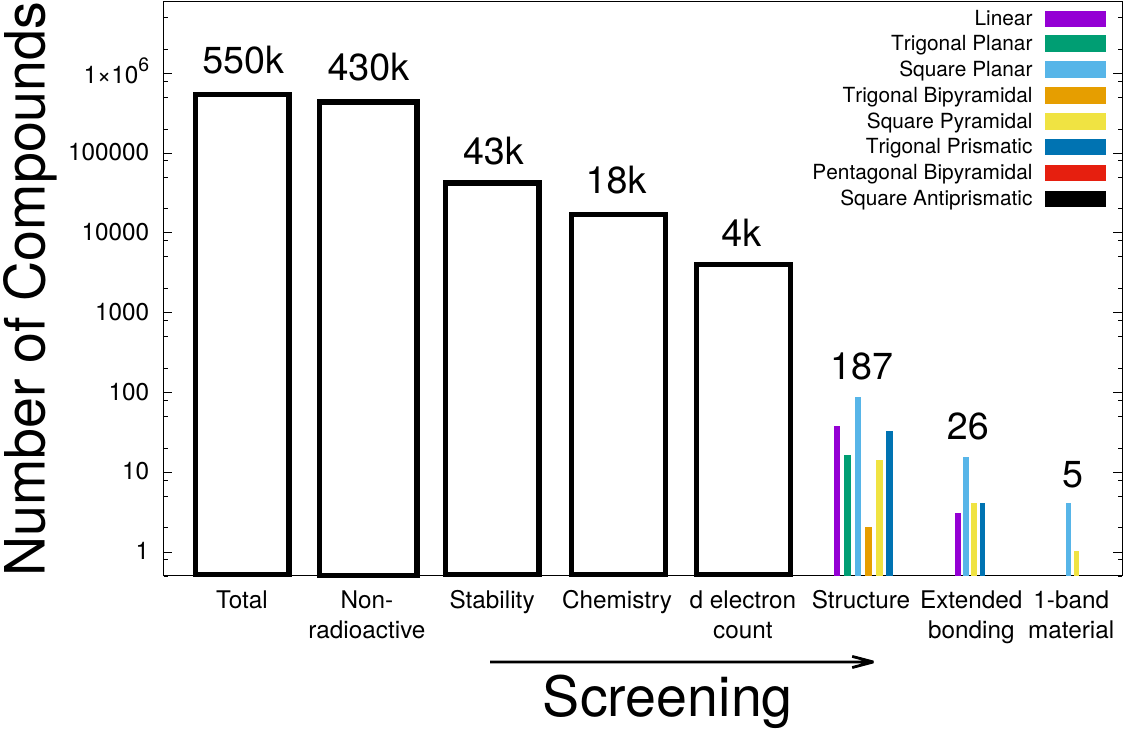}
  \end{center}
  \caption{Materials database screening for correlated one-band
    materials. Decrease in the number of compounds, on a logarithmic
    scale, as additional screening criteria are applied from left to
    right.}
  \label{fig:no_of_compounds}
\end{figure}

We begin by discussing the number of materials obtained via our
high-throughput materials screening strategy. Figure
\ref{fig:no_of_compounds} illustrates how the number of candidate
compounds is reduced from the total as successive filtering criteria
are applied. Note that a logarithmic scale is used since many orders
of magnitude are spanned. Removing compounds containing radioactive
elements eliminates 120,000 compounds from the total of 550,000
candidates. We next filter by stability by including only compounds
which are thermodynamically stable or nearly stable (no more than 25
meV/atom above the ground state). Irrespective of the computed
stability, we also keep any compound if it has been reported
experimentally. 43,000 compounds remain after this filter. 18,000 of
the 43,000 contain a TM and an anion and 4,000 of these have a $d$
electron count compatible with one or more of the desired local
coordination environments. Finally, we find that 187 of the compounds
contain the corresponding coordination environment.

The two biggest decreases (in a fractional sense) of candidate
compounds are (1) the stability filter, in which the number of
compounds is reduced from 430,000 to 43,000 and (2) the local
structure filter, in which the number is reduced from 4,000 to just
187. The substantial decrease in candidate compounds from the
stability filter, a general observation not tied specifically to
searching for one-band correlated materials, is reflective of the high
percentage of unstable, hypothetical compounds in the OQMD derived
from structural prototypes. The significant decrease in candidates
from the structural screening criterion reflects the infrequency of
the low-symmetry coordination environments that lead to
singly-degenerate $d$ orbital levels.

\subsection{Distribution of structure, chemistry, and thermodynamic stability}

\begin{figure}[htbp]
  \begin{center}
    \includegraphics[width=1.0\linewidth]{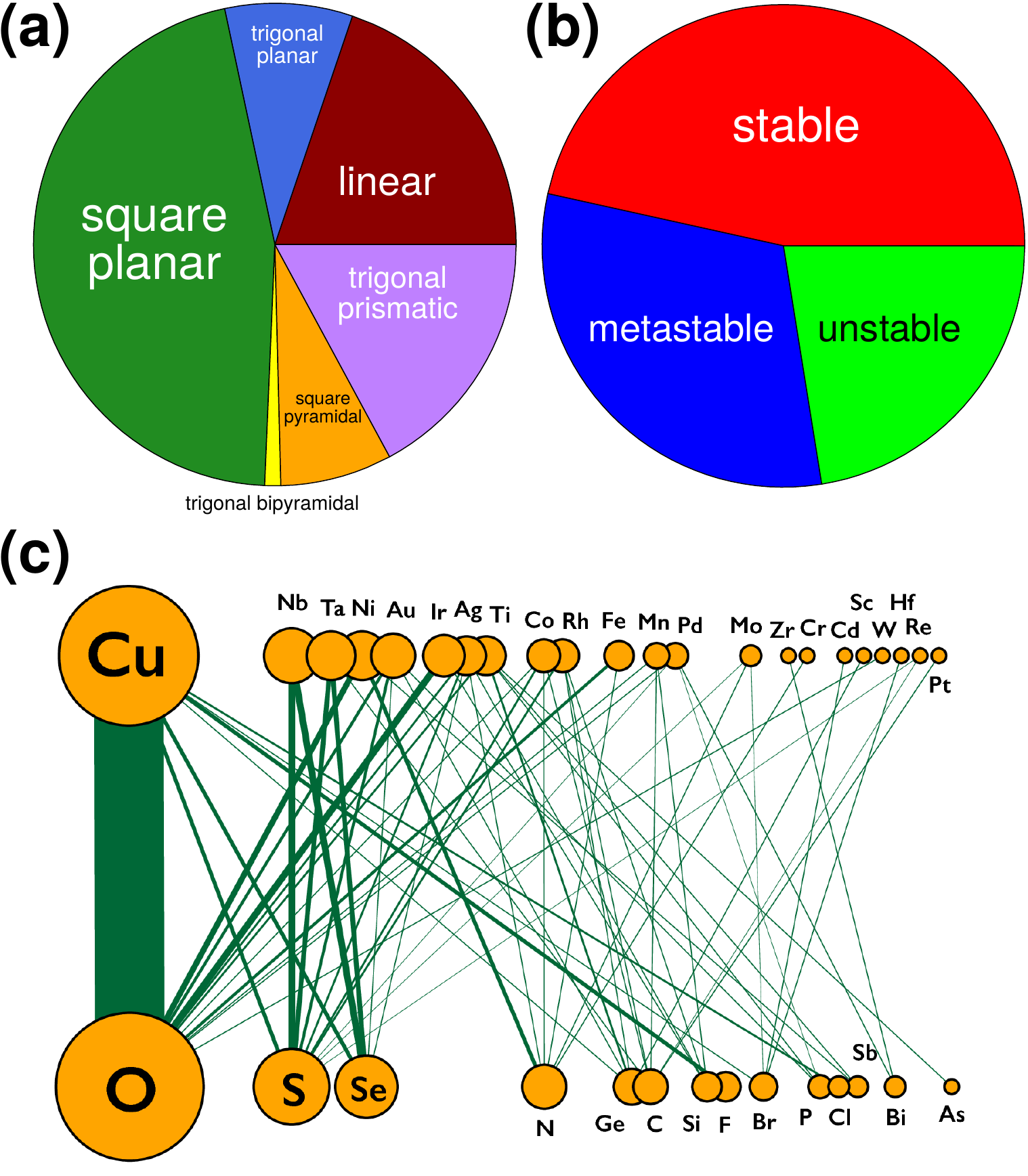}
  \end{center}
  \caption{Characterization of the identified compounds. Distribution
    of the 187 candidate compounds in terms of (a) TM coordination
    geometry, (b) thermodynamic stability, and (c) identity of the TM
    and its nearest-neighbor element.}
  \label{fig:pies_and_graph}
\end{figure}


We now describe the 187 candidate compounds identified by our
screening strategy. The distribution of local TM coordinations for the
187 candidate compounds are displayed in Fig.
\ref{fig:pies_and_graph}(a). Square planar coordination is the most
common, with nearly 50\% of the total. Trigonal prismatic and linear
coordinations are also found in significant numbers. Some of the more
obscure coordinations like pentagonal bipyramidal are not found in a
single compound.

Fig. \ref{fig:pies_and_graph}(b) illustrates the split of
thermodynamic stabilities. Slightly more than half of the compounds
found are thermodynamically stable, and around a third are nearly
stable (no more than 25 meV/atom above the ground state). The
remaining compounds are further than this threshold of stability, but
they have been experimentally observed as they are in the ICSD.

All but 6 of the 187 compounds identified originate from the ICSD. 104
of the 181 ICSD-derived compounds have one of 56 prototypes listed in
the ICSD, which are shown in the Supplemental Material. The most
prevalent ICSD prototypes are trigonal planar La$_3$CuSiS$_7$-type (7,
e.g. Sm$_3$CuSnSe$_7$), trigonal prismatic NbS$_2$-type (5, e.g.
TaSe$_2$), linear CaTiO$_3$-type (5, e.g. InCo$_3$N), and square
planar Ca$_2$CuCl$_2$O$_2$-type (5, e.g. Sr$_2$CoBr$_2$O$_2$).

The only six hypothetical compounds derived from structural prototypes
rather than the ICSD are CsWO$_3$, HfBr$_3$, Mn$_2$TeO, BiPd$_3$,
Ni$_3$Sb, and Zr$_3$Bi. The small number of such compounds suggests
that a small fraction of the structural prototypes used to generate
hypothetical compounds in the OQMD are well-suited to achieve a
one-band correlated material. For example, L$1_2$ is currently the
only OQMD prototype structure for which all sites of an element are in
square planar coordination (e.g. the Cu site in CuAu$_3$). BiPd$_3$,
Ni$_3$Sb, and Zr$_3$Bi correspond to this prototype. Mn$_2$TeO
initially in the $Pnma$ CaFeSeO structure\cite{han_caofese_2015}
relaxes to a distinct $Pnma$ structure with linear Mn--O coordination.
HfBr$_3$ initially in the D0$_{19}$ ($P6_3/mmc$, Ni$_3$Sn-type)
structure relaxes to a distinct $P6_3/mmc$ structure with
one-dimensional chains of triangular-face-sharing trigonal prismatic
HfBr$_6$ units. CsWO$_3$ initially in the $R\bar{3}$ ilmenite
(FeTiO$_3$-type) structure relaxes to a distinct $R\bar{3}$ structure
with layers of edge-sharing trigonal prismatic WO$_6$ units arranged
as a kagome lattice.

The frequency of the TM and anion elements bonded in the low-symmetry
coordinations is illustrated as a network in Fig.
\ref{fig:pies_and_graph}(c). Here each node represents a TM or anion
element and the area of the node (width of the edge) is proportional
to the number of compounds containing these element(s) in the
low-symmetry coordination. Copper-oxygen is the most dominant
chemistry with 74 of the 187 candidate compounds. Cu and O
individually are the most prevalent TM and anion, respectively, as
well: Cu is the TM in 90 of the compounds and O is the anion in 101 of
the compounds. However, there are still many different TMs and anions
represented. The anions S, Se, N and TMs Nb, Ta, Ni, and Au each occur
in at least 9 compounds. The presence of many Cu-containing and square
planar compounds is consistent with a high occurrence of square planar
Cu$^{2+}$ found in a recent statistical analysis of coordination
environments.\cite{waroquiers_statistical_2017}

The complete list of the 187 candidate compounds is included in the
Supplemental Material. We note that the presence of several known
classes of correlated materials such as the cuprates (e.g.
La$_2$CuO$_4$) and group-V transition metal dichalcogenides (e.g.
(Nb/Ta)(S/Se)$_2$) is suggestive of the validity of our screening
strategy.

\subsection{Additional screening criteria for non-cuprates with extended structures}

As discussed above, a key characteristic of strongly correlated
materials like the cuprates is a large $U/t$ ratio. In this regime,
the Coulomb repulsion can overwhelm the electron hopping, leading to
localized electronic states and Mott insulating behavior as is found
in, for example, the cuprate parent compounds. In addition to a large
$U/t$ ratio, a finite hopping $t$ is still necessary to ensure there
is a conduction pathway. As such, we expect materials with extended
crystal structures (leading to extended hopping pathways and
appreciable $t$), to be necessary to achieve a one-band correlated
material.

In order to search for the most promising compounds among the 187
candidates for a one-band correlated material, we therefore perform an
additional post-processing screening criterion based on the crystal
structures to discard any compound for which there is no connectivity
(direct or indirect) between the TM--anion coordination cages. For
example, K$_4$IrO$_4$ is removed since the IrO$_4$ square planar units
are isolated, whereas TlCuPO$_4$ is retained since the CuO$_4$ square
planar units are connected indirectly via phosphate groups.

\begin{table}[htbp]
  \renewcommand{\arraystretch}{0.93}
  \begin{tabular}{|rl|c|}\hline
    \multicolumn{2}{|c|} {Composition} & Coordination \\\hline
    Cs & WO$_3$ & Trig. prismatic \\
    Li & MoN$_2$ & Trig. prismatic\\
    & HfBr$_3$ & Trig. prismatic\\
    Rb & Fe(SeO$_4$)$_2$ & Trig. prismatic\\\hline
    K$_3$H & (CuP$_2$O$_7$)$_2$ & Sq. pyramidal\\
    Al$_2$F$_2$ & CuSi$_2$O$_7$ & Sq. pyramidal\\
    Ba$_4$(Nd/Sm)$_2$ & Cu$_2$O$_9$ & Sq. pyramidal\\\hline
    Ca & NiN & Linear\\
    Sn$_2$ & Co$_3$S$_2$ & Linear\\
    In & Co$_3$N & Linear\\\hline
    (Sr/Ba) & Cu(SeO$_3$)$_2$ & Sq. planar\\
    & CuBr$_2$ & Sq. planar\\
    (La/Gd)$_2$ & Cu(SeO$_3$)$_4$ & Sq. planar\\
    Na$_2$ & CuP$_2$O$_7$ & Sq. planar\\
    Tl & Cu(P/As)O$_4$ & Sq. planar\\
    (Sr/Pb)$_2$ & Cu(BO$_3$)$_2$ & Sq. planar\\
    Bi$_2$ & Cu(SeO$_3$)$_4$ & Sq. planar\\
    Ca$_2$Sb & FeO$_6$ & Sq. planar\\
    Ca(H$_2$O) & CuSiO$_4$ & Sq. planar\\
    Li$_2$ & CuO$_2$ & Sq. planar\\
    Na$_3$ClH & CuPO$_5$ & Sq. planar\\\hline
  \end{tabular}
  \begin{centering}
    \caption{The identified 26 non-cuprate candidate compounds with
      extended electron hopping pathways and the TM coordination
      environment}
    \label{tab:extended_hopping_no_cuprate}
  \end{centering}
\end{table}

Finally, we also remove materials containing any of the CuO$_2$ planes
characteristic of the cuprates since this material class is already
well studied and is not our focus. These two post-processing steps
reduce the number of candidate materials from 187 to 26; these
candidates are contained in Table
\ref{tab:extended_hopping_no_cuprate} and also shown in the ``Extended
bonding'' screening criterion in Fig. \ref{fig:no_of_compounds}. Only
trigonal prismatic, square pyramidal, linear, and square planar
coordinations remain after this additional screening step and
compounds with Cu bonded to O continue to dominate.

\subsection{Electronic band structures}

For each of the 26 compounds, we compute the DFT electronic band
structure to assess whether the underlying one-electron electronic
structure is one-band in nature. In other words, we assess which band
structures contain a single half-filled $d$ band straddling the Fermi
energy with separations in energy below and above the band. The
non-spin-polarized band structure is computed; possible effects of
magnetism and strong electronic correlations on the electronic spectra
will be discussed in the following subsection.

The various screening criteria are not sufficient to \textit{ensure} a
one-band electronic structure, and the majority of the 26 candidates
do not achieve the desired band structure. Several possible
characteristics of the candidate compound can prevent the targeted
band structure:
\begin{enumerate}
\item \textbf{Covalency}: If there is too much hybridization between
  the TM $d$ states and anion $p$ states, an itinerant metal is found
  (example: LiMoN$_2$)
\item \textbf{Metal-metal bonding}: If the crystal structure has
  multiple TM sites and less localized $d$ states (e.g. $5d$), these
  states can hybridize and form a band insulator by completely filling
  a bonding orbital (example: CsWO$_3$)
\item \textbf{Distinct TM sites}: If the crystal structure has TM
  sites with distinct environments, one can find multiple bands
  instead of a single band (example: TlCuPO$_4$)
\end{enumerate} The electronic spectra for these false positive cases is
  included in the Supplemental Material.

\begin{figure*}[htbp]
  \begin{center}
    \includegraphics[width=1.0\linewidth]{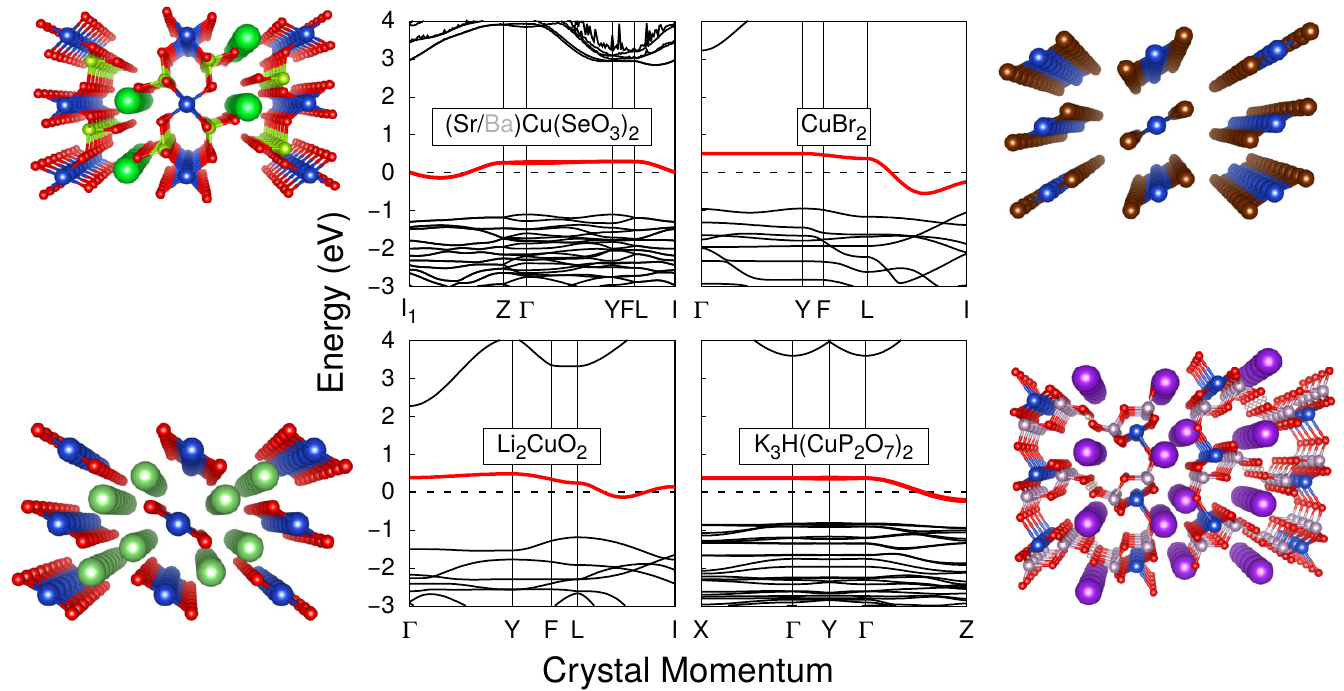}
  \end{center}
  \caption{Electronic band structure of the identified one-band
    correlated materials. Crystal structures and electronic band
    structures within DFT are shown for the five materials. For each
    band structure, there is a single half-filled electronic band. The
    band structure of BaCu(SeO$_3$)$_2$ is not shown since it appears
    nearly identical to that of SrCu(SeO$_3$)$_2$.}
  \label{fig:band_structures}
\end{figure*}

We find 5 materials that successfully achieve a correlated one-band
electronic structure: CuBr$_2$, Li$_2$CuO$_2$, the selenate compounds
BaCu(SeO$_3$)$_2$ and SrCu(SeO$_3$)$_2$, and the pyrophosphate
compound K$_3$H(CuP$_2$O$_7$)$_2$. The crystal structures and
corresponding band structures are depicted in Fig.
\ref{fig:band_structures} and the compounds correspond to the final
``1-band material'' screening criterion in Fig.
\ref{fig:no_of_compounds}. We note that there is a splitting of the
half-filled band for the selenates and pyrophosphate corresponding to
two slightly different TM environments, but the splitting is of very
small magnitude. For example, along the high-symmetry $k$-path for
SrCu(SeO$_3$)$_2$ the maximum splitting (at $\Gamma$) is only 34 meV.

All the compounds contain Cu, suggesting it is difficult to achieve a
one-band correlated material with other TMs. We find that the fraction
of Cu-based candidate materials in Fig. \ref{fig:no_of_compounds}
increases gradually with successive filters, but the most significant
increases occur for the last three (coordination, extended
bonding/non-cuprate, and one-band band structure). Despite the
dominance of oxygen-containing compounds, we do find one compound
lacking oxygen (CuBr$_2$). We also note similarity in the local
structure of the compounds: all but K$_3$H(CuP$_2$O$_7$)$_2$ have
square planar coordination and K$_3$H(CuP$_2$O$_7$)$_2$ is square
pyramidal, which is closely related to square planar.

The selenates and pyrophosphate are all stable or within 1 meV/atom of
the ground state energy. CuBr$_2$ is highly stable: it would have to
increase in energy by 230 meV/atom to become thermodynamically
unstable. Li$_2$CuO$_2$ is 46 meV/atom above the ground state, but it
has been experimentally synthesized. Therefore, all five of the
identified compounds should be ripe for synthesis and further
experimental studies.

While CuBr$_2$ is a binary compound with a simple stoichiometry, we
also find compounds with much more complicated stoichiometries and
structures. While CuBr$_2$ and Li$_2$CuO$_2$ have 1D chains of
edge-sharing Cu--O square planar units, the selenate and pyrophosphate
compounds have more complicated 2D and 3D structures.
K$_3$H(CuP$_2$O$_7$)$_2$ even contains hydrogen. Our query just as
easily finds these more complicated compounds, which is a strength of
the informatics-driven approach to materials discovery.

\subsection{Evidence for strong correlation physics}

\begin{figure}[htbp]
  \begin{center}
    \includegraphics[width=1.0\linewidth]{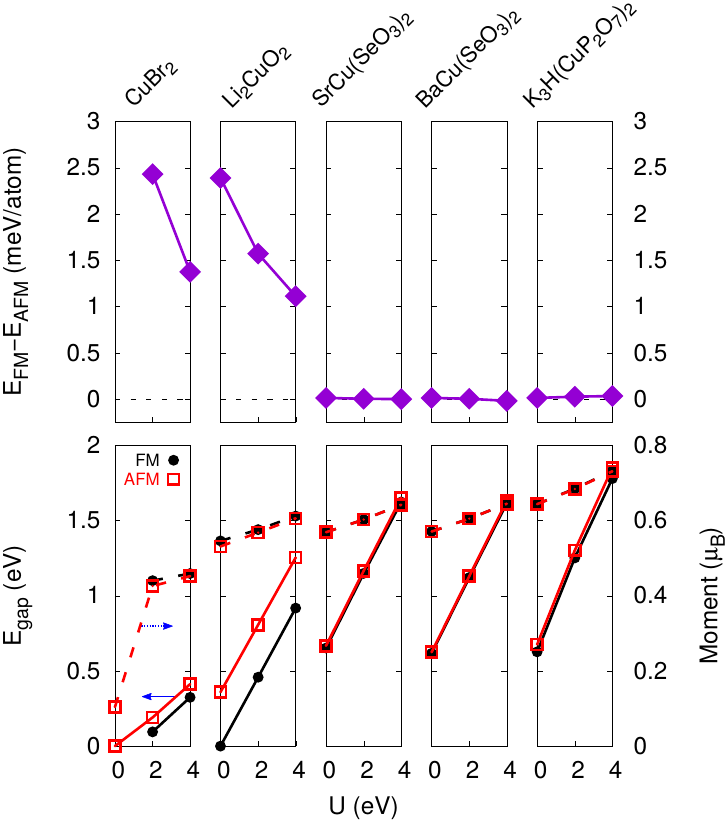}
  \end{center}
  \caption{Impact of electronic correlations via DFT+$U$ on the
    electronic and magnetic properties of the identified materials.
    The relative energetics of ferromagnetic (FM) and
    antiferromagnetic (AFM) states, electronic band gap, and local Cu
    magnetic moment as a function of correlation strength $U$ are
    shown for the 5 candidate materials.}
  \label{fig:dftu}
\end{figure}

Finally, we discuss the possibility of strong correlation physics in
the five identified materials. We note that even at the
non-spin-polarized DFT level of theory, each of these 5 materials
already shows potential for strong correlation physics. In particular,
each has a narrow (bandwidth no more than 1.1 eV) but not completely
flat (bandwidth no less than 0.5 eV) half-filled band, which is
suggestive of a large $U/t$ ratio and finite $t$, of substantial $d$
orbital character. Due to the relatively large size of the primitive
unit cells for (Sr/Ba)Cu(SeO$_3$)$_2$ and K$_3$H(CuP$_2$O$_7$)$_2$, a
quantitative calculation of the $U$ parameter is outside the scope of
this work. However, previous estimations of $U$ for copper compounds
based on \textit{ab initio} calculations and experiments suggest
values will very likely be on the order of 4 eV or
greater.\cite{anisimov_band_1991,liechtenstein_density-functional_1995,galakhov_valence-band_1997,anisimov_first_2002,anisimov_computation_2004,wang_oxidation_2006,nolan_p-type_2006,mazurenko_wannier_2007,onsten_probing_2007,jain_formation_2011,himmetoglu_first-principles_2011,isseroff_importance_2012,ekuma_electronic_2014,mann_first-principles_2016,suarez_structural_2016,brumboiu_influence_2016,cliffe_low-dimensional_2018}

The non-spin-polarized band structure in Fig.
\ref{fig:band_structures} can be considered the \textit{underlying}
one-electron (band theory) electronic structure. Strong electron
correlation can substantially modify the electronic structure in ways
often not adequately captured by DFT, such as magnetism and Mott
insulating behavior. Here we go beyond non-spin-polarized DFT and use
more sophisticated calculations incorporating magnetism and explicit
on-site Coulomb interaction. The goal of such calculations (which
still contain significant approximations) is not to perfectly describe
the electronic properties of the identified materials, but rather to
determine the possibility of strong correlation physics.

In particular, we perform spin-polarized DFT+$U$ calculations on the
five candidate materials. Although DFT+$U$ corresponds to a mean-field
solution to the local correlation problem (exactly solved via
dynamical mean-field theory),\cite{kotliar_electronic_2006} it gives a
baseline expectation of the overall strength of the electronic
correlations at a very cheap computational cost. The results of the
DFT ($U$=0) and DFT+$U$ (finite $U$) calculations including spin
polarization are summarized in Fig. \ref{fig:dftu}, which shows the
dependence of the energetics, band gap, and magnetic moment on the
on-site Coulomb repulsion $U$ for the 5 candidate materials. Several
aspects suggest interesting strong correlation behavior. At the
spin-polarized DFT level, all but CuBr$_2$ exhibit a magnetic
instability corresponding to the formation of local moments. All but
CuBr$_2$ not only become magnetic, but they all fully spin polarize
and open up a band gap (forming a $S=1/2$ state).

DFT+$U$ calculations for $U=2$ and $4$ eV show that for all of these
materials (even the bromide), all of the different magnetic
configurations (ferromagnetic and antiferromagnetic) are gapped. This
suggests Mott insulating behavior in which the strong Coulomb
interaction, rather than a particular magnetic configuration, leads to
an insulating state. In CuBr$_2$ and Li$_2$CuO$_2$, antiferromagnetism
is preferred over ferromagnetism as in the cuprates. In contrast, in
the more complicated crystal structures (the selenates and
pyrophosphate), we find that the ferromagnetic and antiferromagnetic
states are essentially degenerate (less than 1 meV/atom energy
different). This suggests in these materials there is a very weak
magnetic coupling, as might be expected since the square planar and
square pyramidal units are not in close proximity. Since the magnetic
coupling goes as $t^2/U$, the weaker magnetic coupling is consistent
with the smaller bandwidth for these compounds as compared to those of
CuBr$_2$ and Li$_2$CuO$_2$.


The identified materials, whose underlying single-electron band
structure is one-band in nature, exhibit a strong tendency for the
formation of insulating states with fully polarized magnetic moments
regardless of magnetic configuration, as well as the presence of
antiferromagnetic ordering in some cases. While more accurate
correlated calculations and experiments will be necessary to elucidate
the true electronic structure, these results at the approximate
DFT+$U$ level already represent strong evidence for the
possibility of interesting strong correlation physics in these
materials candidates for the rare one-band correlated material.

\section{Conclusions}\label{sec:conclusions}
We employ a materials informatics approach to search for one-band
correlated materials analogous to the cuprate high-temperature
superconductors. Using a query based on transition metal $d$ electron
count, crystal field theory, and formation energy, we search the more
than a half million real and hypothetical inorganic crystals in the
Open Quantum Materials Database for synthesizable materials with an
isolated half-filled $d$ band in the low-energy spectrum. Density
functional theory band structure calculations illustrate that five Cu
compounds, including bromide, oxide, selenate, and pyrophosphate
chemistries, successfully achieve the one-band electronic structure.
Significant evidence of strong correlation physics in the candidate
compounds, including Mott insulating behavior and antiferromagnetism,
is revealed by further calculations including magnetism and on-site
Coulomb interactions. Our data-driven approach opens up new
possibilities for the design and discovery of materials with rare
electronic properties.

\begin{acknowledgments}
We acknowledge support from the U.S. Department of Energy under
Contract DE-SC0015106. Computational resources were provided by the
Quest high performance computing facility at Northwestern University
and the National Energy Research Scientific Computing Center (U.S.
Department of Energy Contract DE-AC02-05CH11231).
\end{acknowledgments}


\bibliography{one_band_materials}

\end{document}